\documentclass[aip,reprint]{revtex4-2}

\usepackage{array}

\usepackage[dvipdfmx]{graphicx}
\usepackage{bmpsize}
\usepackage{color}

\usepackage[utf8]{inputenc}
\usepackage[T1]{fontenc}

\begin{document}

\title{Leveraging Large Language Models and Social Media for Automation in Scanning Probe Microscopy}

\author{Zhuo Diao}
\email[]{enzian0515@gmail.com}
\author{Hayato Yamashita}
\author{Masayuki Abe}
\email[]{abe.masayuki.es@osaka-u.ac.jp}
\affiliation{Graduate School of Engineering Science, Osaka University, 1-3 Machikaneyama, Toyonaka, Osaka 560-0043, Japan}

\begin{abstract}
We present the development of an automated scanning probe microscopy (SPM) measurement system using an advanced large-scale language model (LLM). This SPM system can receive instructions via social networking services (SNS), and the integration of SNS and LLMs enables real-time, language-agnostic control of SPM operations, thereby improving accessibility and efficiency. The integration of LLMs with AI systems with specialized functions brings the realization of self-driving labs closer.
\end{abstract}

\maketitle

As the level of sophistication and reproducibility required for experiments increases, it becomes more difficult to master the experimental equipment. 
This situation implies the need for methods and tools to facilitate experiments. One way to address this problem is through the use of artificial intelligence (AI) and its integration into experimental instruments\cite{onlineai1, onlineai2, onlineai3, onlineai4}.  
In particular, there is widespread interest in creating an environment of automated solutions\cite{sdl1, sdl2}. 
On the other hand, the implementation of fully autonomous systems in experimental research is still in its infancy, despite the promise of revolutionizing research by automating routine tasks and analyses. 
It will still take time for individual researchers to obtain an environment in which they can implement specific experimental AI themselves\cite{aiexp1, aiexp2, aispm1, aispm2, diao2024aiequipped}.
As an alternative trend, research on the use of generative AI\cite{gai_model} in experiments is gaining traction.
Generative AI's capability to create complex data patterns and processes is desirable in increasingly intricate experimental procedures\cite{standarddiff, gai1, gai2, gai3}. 
Among various types of generative AI, large language models (LLMs) stand out for their potential to act as cognitive mediators between researchers' objectives and general tasks. 
Thus, LLMs can be transformative tools, enhancing the efficiency and accessibility of user and experiment instrument interactions.
In this manuscript, we present the use of an advanced LLM in scanning probe microscopy (SPM) experiments, functioning as a measurement agent. 
This SPM system is designed to receive and process instructions via social networking services (SNS), enabling real-time, language-agnostic control of SPM operations. 

\begin{figure}[tb]
\begin{center}
\includegraphics[width=85mm]{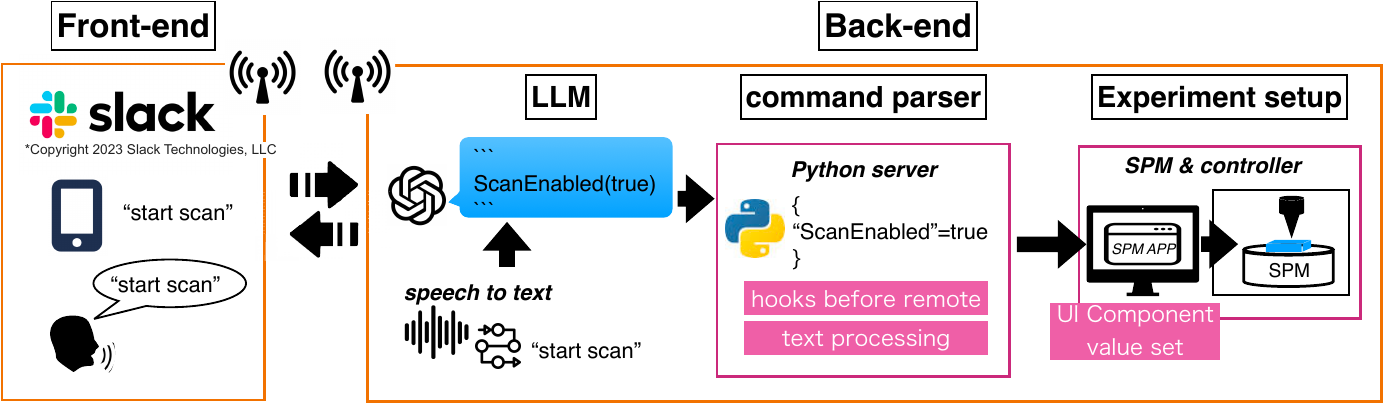}
\caption{\label{SNSschematic}
Schematic diagram of interactive system of scanning probe microscopy (SPM). 
In the experiments, Slack, a social networking service (SNS), is used for the human input portion of the commands, which ChapGPT, a large-scale language model (LLM), understands and converts into actual commands for the SPM device.
} 
\end{center}
\end{figure}

Figure \ref{SNSschematic} shows an SPM system integrated with LLM and SNS.
Our system integrates a front-end SNS client utilizing Slack web app\cite{slack} for receiving user commands, and a back-end embedded system for command processing and execution. 
To deal with user actions, we utilize Slack API's event subscription feature, allowing the system to listen to the user messages that are sent to specific Slack channels.
User messages can be in multiple languages, and different message types (text/audio).
The SPM agent involves translating user commands into executable SPM commands via a sophisticated AI model that processes natural language and speech inputs, leveraging OpenAI's Whisper\cite{whisper} for speech-to-text (TTS) conversion and ChatGPT (gpt-4-0124-preview)\cite{gpt4} for understanding and generating appropriate commands for SPM control (text-to-command: TTC).
The system's core, a Python server, hosts the custom SPM controller, facilitating direct TCP connection with the SPM instrument. 
The SPM controller can receive commands in JSON format to set the value of the corresponding UI element name.
In Fig.~\ref{SNSschematic}, as an example, for a switchable UI named ``ScanEnabled", giving it a true value will control the SPM to start scanning. 

Constructing clear and detailed LLM prompts regarding parameter settings and behavior in the experiment is important for the LLM to make the following decisions from the incoming messages.
The following prompt is an example of a configuration in our system.

\vspace{5mm}

\begin{quote}
\itshape
``You are a robot controlling a scanning probe microscopy. Users will provide instructions in text format on how to control the device, and you'll need to translate these texts into specific programmatic commands. " ,
``You can control and set the scan parameters in the scanning probe microscopy.  " ,
 ``The below list shows the function about what you can do. Each programmatic command is followed by parentheses containing the names of required parameters. Following this is a description of the command, arg type indicates the type of parameter, and arg description represents the parameter's description." + command str, 
``You should try your best to understand the instructions and use the list up functions to write. The function argument should follow the type I defined.",
``If the user's instructions can be accomplished by multiple step commands, then output them sequentially and separate each command with a new line.",
 ``If the user's instructions cannot be carried out by the commands provided above alone, please respond with 'None' first and then give a reason to user. Otherwise, reply with the names of the corresponding programmatic commands and provide appropriate values within parentheses. ", 
``Reply to me in language str."
\end{quote}
\vspace{5mm}
In the experiment, ChatGPT determines the instructions sent to SPM based on these prompts as well as five previous message histories and the current user message.
``language str" will be replaced with the detected language name of the user message.
This ensures that our measurement agent replies to the user in the same language they used.
Table \ref{tab:commands} shows the SPM experiment commands currently understood by the SPM agent. 
These commands are saved in a bulleted format, such as a CSV file so that they can be easily added or deleted.
The bulleted command list we used in this paper is shown in Tab. \ref{tab:commands}. 
In the above prompts, "``command str" will be replaced with the text content of this bulleted list.
The table serves as an API reference documentation for the LLM, informing the supported SPM control command and explaining the usage of the commands and their arguments.

\begin{table}[bt]  
\begin{center}
\renewcommand{\arraystretch}{1}  
\caption{Description of Programmatic Commands for ChatGPT. Descriptions in the table is input to ``command str" in the prompt.
\label{tab:commands} }
{\scriptsize
\begin{tabular}{
|>{\raggedright\arraybackslash}p{2.0cm}|>{\raggedright\arraybackslash}p{2.1cm}|>{\raggedright\arraybackslash}p{0.5cm}|>{\raggedright\arraybackslash}p{2.1cm}|}
\hline
\textbf{Programmatic Commands} & \textbf{Description} & \textbf{Arg Type} & \textbf{Arg Description} \\
\hline
StageOffset\_X\_Tube (arg) & Setting the absolute  X (left and right direction) position coordinates of the probe (SPM tip) & float & Coordinate value of X in nanometers \\
\hline
StageOffset\_Y\_Tube (arg) & Setting the absolute  Y (up and down direction) position coordinates of the probe (SPM tip) & float & Coordinate value of Y in nanometers. \\
\hline
StageOffset\_X\_Tube\_ ADD (arg) & Sets the relative X (left/right direction) position to be moved from the current coordinates of the probe (SPM tip) & float & Distance to be move in  X, in nanometers \\
\hline
StageOffset\_Y\_Tube\_ ADD (arg) & Sets the relative Y (up/down direction) position to be moved from the current coordinates of the probe (SPM tip) & float & Distance to be move in Y, in nanometers \\
\hline
Sample\_Bias (arg) & The bias voltage add to the measured sample & float & the value of the sample bias \\
\hline
Aux1MaxVoltage (arg) & Setting the scanning area range of the X size & float & scan range of the X direction in nanometers \\
\hline
Aux2MaxVoltage (arg) & Setting the scanning area range of the Y size & float & scan range of the Y direction in nanometers \\
\hline
ScanEnabled (arg) & Switches to control scanning & bool & true to start the scan and false to stop the scan \\
\hline
\end{tabular}
}
\end{center}
\end{table}

The output text generated by ChatGPT is then processed via a command parser, which converts it into a canonical format of the SPM command.
Given the inherent uncertainty in AI-generated output, the command parser ensures that the resulting command line for controlling the SPM device adheres to the system's specifications.
Unsupported output content is disregarded, while supported programmatic commands and their corresponding arguments are extracted.

Using the above-mentioned system, we perform the STM experiment operated in ultra-high vacuum condition ($< 1 \times 10^{-8}\  \mathrm{Pa}$) at room temperature.
Mechanically etched PtIr STM tips (UNISON P-50 Pt; Ir) were used for imaging. 
As a sample, we prepare $\mathrm{Si}(111)$ clean surface by a standard flashing-and-annealing method. 
For data acquisition, we have established an SPM system built with LabVIEW, LabVIEW FPGA, and Python.
NI PXIe-7857R was used as the measurement board, 
The scanning and data acquisition were controlled with Python script in which automated SPM measurement routines and scan functions for optimizing experimental environment were contained.

\begin{figure}[tbh]
\begin{center}
\includegraphics[width=85mm]{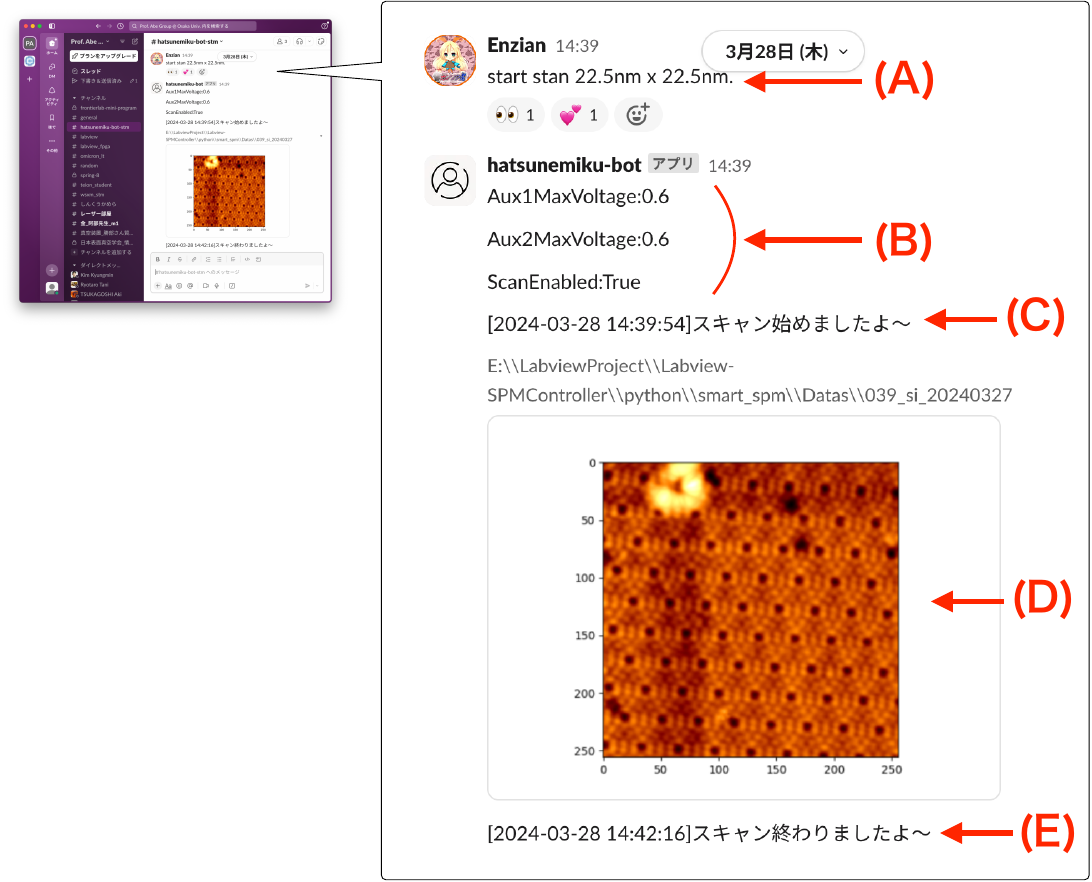}
\caption{\label{fig:stm1}
An example of an STM experiment using SNS. 
}
\end{center}
\end{figure}

Figure \ref{fig:stm1} shows a demonstration of STM measurements of the $\mathrm{Si(111)-(7 \times 7)}$ surface using SNS.
In part (A) of the figure, the operator inputs the incomplete English phrase "start scan 22.5 nm x 22.5 nm," intending to mean ``Start scanning in the area of $\mathrm{22.5 nm \times 22.5 nm}$".
The backend system interprets the messages and sends the command (as shown in (B)) to the SPM.
Here, $\mathrm{0.6 V}$ is the calibrated voltage value of the scanner equivalent to $\mathrm{22.5 nm}$.
Thus, even if the English are not perfect, the correct command is sent, thanks to LLM.
Part (C) and (E) show that the SPM agent sends information messages indicating that scanning has started and finished, respectively.
After the measurement, the scanned image is sent back to the operator as shown in (D). The image is filtered for easier viewing.

\begin{figure}[tbh]
\begin{center}
\includegraphics[width=85mm]{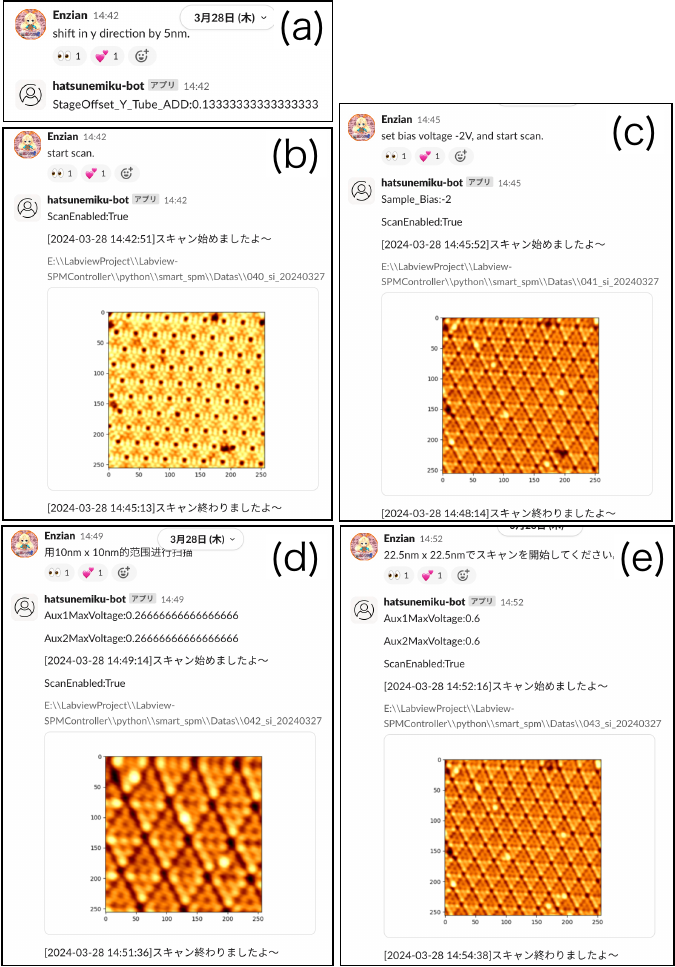}
\caption{\label{fig:stm2}
The messages for image acquisition and responses in succession.
}
\end{center}
\end{figure}

Figure \ref{fig:stm2} shows the results of the experiment following those presented in Fig. \ref{fig:stm1}.
As the STM image from Fig. \ref{fig:stm1} depicted an adsorbed object in the upper left portion, to avoid the ares, a message  "shift in the y-direction by 5 nm" was entered as shown in Fig. \ref{fig:stm1} (a), then the tip positioning update was performed by $\mathrm{5\ nm}$.
Figure \ref{fig:stm2}(b) confirms that the scanning avoided the adsorbed object by restarting the scan.
In Figure 3(c), a message "set bias voltage to -2V and start scan" was input, setting the bias voltage to $\mathrm{-2\ V}$ and beginning the scanning of $\mathrm{22.5\ nm \times 22.5\ nm}$ area.
Since this negative bias was applied, both faulted and unfaulted half-unit cells are distinguishable. 
These steps illustrate that the two commands (changing the bias voltage and initiating scanning) were executed correctly.
Although the scanning area was not explicitly mentioned in Fig.\ref{fig:stm2}(c), ChatGPT presumed the previously used scanning range as acceptable based on the command history and proceeded with the scan.
In Fig.\ref{fig:stm2}(d), a message meaning ``Start scanning in the area of $\mathrm{10 nm \times 10 nm}$" was entered in Chinese. 
In the STM images on Slack, the start position of scanning is the upper left.
Comparing with Fig. \ref{fig:stm2}(c), we can see that the $\mathrm{10\ nm \times 10\ nm}$ area in the upper left portion is imaged.
Additionally, Figure \ref{fig:stm2}(e) shows a message entered in Japanese to scan a $\mathrm{22.5\ nm \times 22.5\ nm}$ area, resulting in the same image as seen in Fig.\ref{fig:stm2}(c). 
This demonstrates that, although the ChatGPT prompt is written in English, the LLM can interpret and execute commands in multiple languages.

\begin{figure}[tbh]
\begin{center}
\includegraphics[width=55mm]{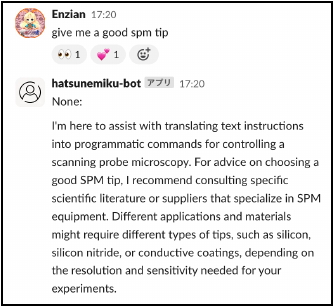}
\caption{\label{fig:stm3}
Response from the SPM agent when a message related to a command that has not been preconfigured is sent.
}
\end{center}
\end{figure}

According to the prompts, when ChatGPT judges the user message as "not executable", our system will extract the reason part from the generated text and send it to the Slack client. 
Figure \ref{fig:stm3} shows a result when a command not in Tab. \ref{tab:commands}  is entered into the message; the prompt instructs the user to answer "None" if the command cannot be executed and to tell the user why, and the message that shows the reason will be generated. 

This study has presented the stage for a detailed discussion of the integration of AI into the experimental process, exploring both its potential benefits and challenges. 
One of the primary goals of AI in scientific experiments is to realize a self-driving lab: to enable AI to perform preset experiments autonomously.
On the other hand, depending on the research (for example imaging with microscopy), it will be necessary to check the results of the experiments in real-time. 
In such cases, humans will intervene in the experiment, but even then, they will always be constrained by the time it may take to check reproducibility and the experiment itself. 
SPM often requires not only skilled experimental techniques but also a very long time for data acquisition. In particular, while in spectroscopic measurements and atom manipulation experiments, it is essential to set up areas and parameters that allow appropriate data acquisition\cite{Sugimoto:2005aa, Sugimoto:2007Nature}. 
In such experimental studies, adding human guidance is an effective way to perform effective experiments and provide feedback to train AI systems capable of autonomously performing experiments.
SPM operators have been dominated by manual and often labor-intensive processes\cite{drift2, tipfix1, Extance:2018aa}. 
As the complexity of experiments grows, so too does the challenge of mastering intricate methodologies, particularly for novices in the field. 
This underscores a critical need for innovative solutions that can streamline the experimental process and make high-quality research more accessible and reproducible.

The use of SNS and LLM proposed in this study is one way to solve these problems.
As a prospect, LLM is expected to be utilized in a wide range of experimental applications. 
A general-purpose LLM such as ChatGPT could serve as the "brain of an experimental system" coordinating with other specialized algorithms and AI models to handle complex, specialized research tasks.
The specialized AI model can extract features from the data in a higher-dimensional space compared to the model output\cite{simonyan2015deep, NIPS2012_c399862d}(e.g., classifying a 256$\times$256 pixel SPM image into a few labels of feature data, which will make LLM easier to understand).
This feature information can then be incorporated into user prompts, allowing the LLM to determine the next experimental procedure using predefined prompts.
Then, LLM can perform efficient experiments using more advanced SPM features, after we implement these methods into the system and enrich the method description in the document file of Tab.~\ref{tab:commands}. 
For example, to perform the automated SPM experiments, several advanced algorithms and AI implementations for SPM can be integrated into the system, including data processing methods\cite{spmprocessing, spmprocessing1}, drift correction\cite{drift}, tip reconstruction\cite{tipfix, aispm2}, and atom manipulation\cite{am}.

%
The accuracy of the AI can be improved by generating more detailed prompts, similar to instruction manuals.
Fortunately, the ``gpt-4-0125-preview" model supports up to 128,000 tokens, and given that a single conversation typically utilizes only around 500 tokens, the full potential of ChatGPT is not yet being leveraged. 
The AI can facilitate more advanced experimental operations by incorporating detailed instructions in the experimental apparatus into the prompts.

In summary, we have developed an SPM agent representing an innovative application of AI in scientific research, offering a user-friendly and efficient method for conducting SPM experiments remotely. This work contributes to the broader efforts in automating scientific research, promising to accelerate experimental workflows and facilitate remote experimentation in various scientific fields.

This work was supported in-part by a Grant-in-Aid for Scientific Research (19H05789, 21H01812, 22K18945) from the Ministry of Education, Culture, Sports, Science and Technology of Japan (MEXT).

\bibliography{ref}

\providecommand{\newblock}{}
\begin{thebibliography}{10}
\expandafter\ifx\csname url\endcsname\relax
  \def\url#1{{\tt #1}}\fi
\expandafter\ifx\csname urlprefix\endcsname\relax\def\urlprefix{URL }\fi
\providecommand{\eprint}[2][]{\url{#2}}

\bibitem{onlineai1}
Coley C~W, Thomas D~A, Lummiss J~A~M, Jaworski J~N, Breen C~P, Schultz V, Hart
  T, Fishman J~S, Rogers L, Gao H, Hicklin R~W, Plehiers P~P, Byington J,
  Piotti J~S, Green W~H, Hart A~J, Jamison T~F and Jensen K~F 2019 {\em
  Science\/} {\bf 365} eaax1566

\bibitem{onlineai2}
Kaminski T~S and Garstecki P 2017 {\em Chem. Soc. Rev.\/} {\bf 46}(20)
  6210--6226

\bibitem{onlineai3}
Epps R~W, Bowen M~S, Volk A~A, Abdel-Latif K, Han S, Reyes K~G, Amassian A and
  Abolhasani M 2020 {\em Adv Mater\/} {\bf 32} e2001626 ISSN 1521-4095
  (Electronic); 0935-9648 (Linking)

\bibitem{onlineai4}
Tao H, Wu T, Kheiri S, Aldeghi M, Aspuru-Guzik A and Kumacheva E 2021 {\em
  Advanced Functional Materials\/} {\bf 31} 2106725

\bibitem{sdl1}
Abolhasani M and Kumacheva E 2023 {\em Nature Synthesis\/} {\bf 2} 483--492

\bibitem{sdl2}
Rooney M~B, MacLeod B~P, Oldford R, Thompson Z~J, White K~L, Tungjunyatham J,
  Stankiewicz B~J and Berlinguette C~P 2022 {\em Digital Discovery\/} {\bf
  1}(4) 382--389

\bibitem{aiexp1}
Gromski P~S, Henson A~B, Granda J~M and Cronin L 2019 {\em Nature Reviews
  Chemistry\/} {\bf 3} 119--128

\bibitem{aiexp2}
Epps R~W and Abolhasani M 2021 {\em Applied Physics Reviews\/} {\bf 8} 041316

\bibitem{aispm1}
Vasudevan R~K, Kelley K~P, Hinkle J, Funakubo H, Jesse S, Kalinin S~V and
  Ziatdinov M 2021 {\em ACS Nano\/} {\bf 15} 11253--11262

\bibitem{aispm2}
Krull A, Hirsch P, Rother C, Schiffrin A and Krull C 2020 {\em Communications
  Physics\/} {\bf 3} 54

\bibitem{diao2024aiequipped}
Diao Z, Ueda K, Hou L, Li F, Yamashita H and Abe M 2024 Ai-equipped scanning
  probe microscopy for autonomous site-specific atomic-level characterization
  at room temperature (\textit{Preprint} \eprint{2404.11162})

\bibitem{gai_model}
Granovetter M 1978 {\em American Journal of Sociology\/} {\bf 83} 1420--1443

\bibitem{standarddiff}
Rombach R, Blattmann A, Lorenz D, Esser P and Ommer B 2021 High-resolution
  image synthesis with latent diffusion models (\textit{Preprint}
  \eprint{2112.10752})

\bibitem{gai1}
Hoogeboom E, Satorras V~G, Vignac C and Welling M 2022 Equivariant diffusion
  for molecule generation in 3d vol 162 pp 8867 -- 8887 cited by: 82

\bibitem{gai2}
Sanchez-Lengeling B and Aspuru-Guzik A 2018 {\em Science\/} {\bf 361} 360--365

\bibitem{gai3}
Zhao Y, Siriwardane E~M~D, Wu Z, Fu N, Al-Fahdi M, Hu M and Hu J 2023 {\em npj
  Computational Materials\/} {\bf 9} 38

\bibitem{slack}
Slack is a trademark and service mark of Slack Technologies, Inc., registered
  in the U.S. and in other countries.

\bibitem{whisper}
Radford A, Kim J~W, Xu T, Brockman G, McLeavey C and Sutskever I 2022 Robust
  speech recognition via large-scale weak supervision (\textit{Preprint}
  \eprint{2212.04356})

\bibitem{gpt4}
OpenAI, Achiam J, Adler S, Agarwal S and Ahmad L 2024 Gpt-4 technical report
  (\textit{Preprint} \eprint{2303.08774})

\bibitem{Sugimoto:2005aa}
Sugimoto Y, Abe M, Hirayama S, Oyabu N, Custance {\'O} and Morita S 2005 {\em
  Nature Materials\/} {\bf 4} 156--159

\bibitem{Sugimoto:2007Nature}
Sugimoto Y, Pou P, Abe M, Jelinek P, P{\' e}rez R, Morita S and Custance {\'O}
  2007 {\em Nature\/} {\bf 446} 64--67

\bibitem{drift2}
Abe M, Sugimoto Y, Custance {\'O} and Morita S 2005 {\em Nanotechnology\/} {\bf
  16} 3029

\bibitem{tipfix1}
Paul W, Oliver D, Miyahara Y and Gr{\"u}tter P 2014 {\em Applied Surface
  Science\/} {\bf 305} 124--132

\bibitem{Extance:2018aa}
Extance A 2018 {\em Nature\/} {\bf 555} 545--547 ISSN 1476-4687 (Electronic);
  0028-0836 (Linking)

\bibitem{simonyan2015deep}
Simonyan K and Zisserman A 2015 Very deep convolutional networks for
  large-scale image recognition (\textit{Preprint} \eprint{1409.1556})

\bibitem{NIPS2012_c399862d}
Krizhevsky A, Sutskever I and Hinton G~E 2012 Imagenet classification with deep
  convolutional neural networks {\em Advances in Neural Information Processing
  Systems\/} vol~25 ed Pereira F, Burges C, Bottou L and Weinberger K (Curran
  Associates, Inc.)

\bibitem{spmprocessing}
Jones L, Yang H, Pennycook T~J, Marshall M~S~J, Van~Aert S, Browning N~D,
  Castell M~R and Nellist P~D 2015 {\em Advanced Structural and Chemical
  Imaging\/} {\bf 1} 8

\bibitem{spmprocessing1}
Diao Z, Katsube D, Yamashita H, Sugimoto Y, Custance O and Abe M 2020 {\em
  Applied Physics Letters\/} {\bf 117} 033104

\bibitem{drift}
Diao Z, Ueda K, Hou L, Yamashita H, Custance O and Abe M 2023 {\em Applied
  Physics Letters\/} {\bf 122} 121601

\bibitem{tipfix}
Diao Z, Hou L and Abe M 2023 {\em Applied Physics Express\/} {\bf 16} 085002

\bibitem{am}
Chen I~J, Aapro M, Kipnis A, Ilin A, Liljeroth P and Foster A~S 2022 {\em
  Nature Communications\/} {\bf 13} 7499

\end{thebibliography}

\end{document}